\documentclass[a4paper,doc,natbib]{apa7}

\usepackage{amsmath,amssymb,amsfonts,amsthm}
\usepackage{graphicx}
\usepackage{natbib}
\usepackage{nicefrac}
\usepackage{float}
\usepackage{booktabs}

\usepackage[ruled,vlined]{algorithm2e}

\usepackage{enumitem}
\setlist{nolistsep}

\usepackage[dvipsnames]{xcolor}

\newcommand{\prob}[1]{\mathbb{P}\left({#1}\right)}
\newcommand{\likelihoodsym}{\mathbb{L}}
\newcommand{\likelihood}[1]{\likelihoodsym\left({#1}\right)}
\newcommand{\prior}[1]{\pi\left({#1}\right)}

\newcommand{\dnorm}[1]{\mathrm{Normal}\left({#1}\right)}

\newcommand{\predictor}{\beta}

\newcommand{\indicator}{\gamma}
\newcommand{\indicators}{\bm{\indicator}}

\newcommand{\reals}{\mathbb{R}}
\newcommand{\parameter}{\theta}
\newcommand{\parameters}{\bm{\parameter}}
\newcommand{\velocity}{v}
\newcommand{\velocities}{\bm{\velocity}}

\title{Accelerating Bayesian Variable Selection using Piecewise Deterministic Markov Processes}

\authorsnames[1,2]{D. {van den Bergh}, M. Marsman}
\authorsaffiliations{
    {Tilburg School of Social and Behavioral Sciences: Department of Methodology and Statistics, Tilburg University},
    {Department of Psychology, University of Amsterdam}
}

\shorttitle{Accelerating Bayesian Variable Selection using PDMPs}
\abstract{
Bayesian variable selection becomes computationally challenging when models contain many dependent parameters. We study Piecewise Deterministic Markov Process (PDMP) samplers as a continuous-time alternative to conventional Markov chain Monte Carlo for spike-and-slab variable selection. In sticky PDMP samplers, active parameters evolve continuously until they reach zero, where they may remain for a random duration before re-entering the model. While one parameter enters or leaves the model, the remaining parameters continue to evolve along the deterministic flow, offering a potentially advantageous mechanism for exploring posteriors with strongly dependent parameters. We make two methodological contributions. First, we extend existing sticky PDMP methods beyond independent spike-and-slab priors to dependent model priors and dependent slab distributions. Second, we investigate the use of unbiased stochastic gradients to reduce the computational cost of variable selection when the likelihood decomposes into many factors while retaining the same target distribution. We study these extensions to two psychometric models: a Gaussian random intercept cross-lagged panel model and an ordinal Markov random field. For the former, marginalization yields a fixed-dimensional sufficient-statistic representation that permits efficient model evaluation. For the latter, the model factorizes, which enables subsampling over person-node contributions. In simulation studies, we compare ZigZag, Bouncy Particle, and Boomerang dynamics with reversible-jump MCMC, and examine the effects of prior dependence and stochastic-gradient subsampling on sampling efficiency. We illustrate the methodology using data from an empirical study on mental well-being. Finally, we discuss the advantages and challenges when using PDMP samplers for Bayesian variable selection.
}

\begin{document}

\maketitle

\section{Introduction}

Variable selection is a fundamental task in statistical modeling.
In the Bayesian framework, variable selection is typically approached by placing spike-and-slab priors on the coefficients of interest \citep{mitchell1988bayesian, GeorgeMcCulloch_1993}.
These priors consist of a point mass at zero, the spike, and a continuous distribution, the slab, which together induce a joint posterior distribution over models and parameters.
The most common computational strategies for exploring this posterior are Markov chain Monte Carlo (MCMC) methods based on stochastic search variable selection \citep{GeorgeMcCulloch_1993} or reversible jump MCMC \citep{Green_1995}.
However, these traditional MCMC approaches can become computationally expensive when the number of parameters is large \citep{ishwaran2005spike, TadeseVanucci_2022_HanbookVarSel}.

In psychometrics, variable selection is simultaneously of substantive interest and a major computational challenge.
For instance, in network psychometrics, edges in the network represent conditional associations between variables, and selecting which edges to include is important for obtaining an interpretable network structure.
However, the number of possible edges grows quadratically with the number of variables, complicating the selection problem.
Similarly, in longitudinal panel models such as the random intercept cross-lagged panel model \citep{hamaker2015critique}, researchers are often interested in determining which cross-lagged effects are nonzero, and the number of such effects likewise grows quadratically with the number of variables.

A recent alternative to traditional MCMC for Bayesian inference is provided by Piecewise Deterministic Markov Processes (PDMPs; \citealp{davis1993markov, fearnhead2018piecewise}).
In contrast to traditional MCMC, where iterations occur in discrete time, PDMPs are continuous-time stochastic processes that move according to deterministic dynamics between random event times.
At an event, the state changes instantaneously and the process continues along a new deterministic trajectory.
The event rates and transitions are constructed so that the target posterior distribution is stationary for the resulting process.
Several concrete PDMP samplers have been developed, including the Zig--Zag sampler \citep{bierkens2019ZigZag}, the Bouncy Particle sampler \citep{bouchard2018Bouncy}, and the Boomerang sampler \citep{bierkens2020boomerang}.

To accommodate variable selection, \citet{bierkens2023sticky} introduced sticky events, which allow parameters to ``stick'' to zero for random durations.
When combined with a PDMP sampler, sticky events provide a continuous-time implementation of spike-and-slab variable selection.
An active parameter becomes excluded when its deterministic trajectory reaches zero, while the remaining parameters continue to evolve.
Conversely, an excluded parameter remains fixed at zero until an unfreezing event returns it to the continuous part of the state space.
This differs from common reversible-jump implementations based on local model-space proposals, where the inclusion status of one or a small number of parameters is changed at a time while the remaining parameters are left unchanged \citep{Green_1995}.
Such local proposals can become inefficient when posterior dependence between parameters is strong, because coordinated changes may be required to move efficiently between models.
Another advantageous property of PDMPs is that they can use unbiased stochastic gradient estimates while retaining the exact target distribution \citep{bierkens2019ZigZag, pakman2017stochastic}.
In practice, this can reduce computational cost when the likelihood decomposes into many factors and the corresponding event-time construction can be simulated without evaluating the full gradient.

This paper makes two methodological contributions to PDMP-based variable selection for psychometric models.
First, previous work on sticky PDMP methods focused on independent spike-and-slab priors \citep{bierkens2023sticky}.
We extend the sticky construction to dependent priors, allowing both dependence among inclusion indicators through model-size priors and dependence among included coefficients through joint slab distributions.
This includes Beta--Bernoulli model priors, which induce dependence among inclusion indicators and provide multiplicity adjustment \citep{ScottBerger_2010}, as well as Gaussian scale-mixture slabs with shared scale parameters.
Second, we investigate the use of unbiased stochastic gradients to reduce the computational cost of PDMP variable selection when the likelihood admits a suitable factorization.
We evaluate the proposed methods on two psychometric models in a simulation study.
The first is the random intercept cross-lagged panel model (RI-CLPM; \citealp{hamaker2015critique, mulder2021three}), a longitudinal model with a potentially large number of cross-lagged effects.
The second is the ordinal Markov random field (OMRF; \citealp{MarsmanEtAl_2023_ordinal}), a network model for ordinal data with a potentially large number of pairwise interactions.
We compare three PDMP dynamics, the Zig--Zag sampler, the Bouncy Particle sampler, and the Boomerang sampler, against a reversible-jump sampler for the same models implemented in NIMBLE \citep{nimble-article:2017}.
% Afterward, we apply the RI-CLPM to an empirical four-wave panel dataset on well-being to illustrate the workflow in an applied setting.
Afterward, we apply the OMRF to an empirical dataset on mental well-being.
We conclude by discussing the implications and limitations of the proposed PDMP variable selection methods.

\section{Piecewise Deterministic Markov Processes}

Piecewise Deterministic Markov Processes are a class of continuous-time stochastic processes characterized by deterministic motion interrupted by random jumps \citep{davis1993markov}.
In the context of Bayesian inference, PDMPs are used to construct Markov processes that have the target posterior distribution as their stationary distribution \citep{fearnhead2018piecewise}.
A PDMP is defined on an augmented state space that includes both the parameter vector $\parameters \in \reals^d$ and a velocity vector $\velocities \in \mathbb{V} \subseteq \reals^d$.
The process consists of deterministic dynamics, an event rate, and a transition kernel at events.

\paragraph{Deterministic Dynamics}
The deterministic dynamics are specified through an ordinary differential equation $\frac{\partial (\parameters, \velocities)}{\partial t} = \phi(\parameters, \velocities)$, which implies that there is a deterministic transition function.
If the current time is $t$ and $\Delta t$ elapses, a function $\Phi$ maps the current state to the new state:
\begin{align}
\parameters(t+\Delta t), \velocities(t+\Delta t) = \Phi(\parameters(t), \velocities(t), \Delta t).
\end{align}
The time index is continuous rather than a sequence of equally spaced MCMC iterations, hence this is a continuous-time process.

\paragraph{Event Rate}
Random events interrupt the deterministic flow.
The event rate $\lambda(\parameters, \velocities)$ determines their distribution.
The time to the next event depends on the integrated rate along the deterministic path.
Its survival function is
\begin{align}
P(\Delta t > s \mid \parameters(t), \velocities(t)) = \exp\left(-\int_0^s \lambda(\parameters(t+u), \velocities(t+u)) \, du\right)
\end{align}
When developing new PDMP dynamics, the event rate is typically designed so that the resulting process has the target posterior as its stationary distribution \citep{fearnhead2018piecewise}.

\paragraph{Transition Distribution}
When an event occurs at time $t' = t+\Delta t$, the state transitions instantly from $(\parameters(t'), \velocities(t'))$ to a new state $(\parameters'(t'), \velocities'(t'))$.
Typically this only changes the velocity.

We now describe three concrete samplers: the Zig--Zag sampler, the Bouncy Particle sampler, and the Boomerang sampler.
For a detailed introduction on PDMPs, see \citet{fearnhead2018piecewise, bierkens2019ZigZag} and for a measure-theoretic foundation, see \citet{davis1993markov}.

\subsection{Zig--Zag Sampler}

The Zig--Zag sampler \citep{bierkens2019ZigZag} is the simplest of the three PDMP samplers we study.
Its velocity vector is restricted to $\velocities \in \{-1, +1\}^d$, meaning each parameter moves at constant speed in either the positive or negative direction.
The deterministic dynamics are simply
\begin{align}
\parameters(t + \Delta t) = \parameters(t) + \velocities \, \Delta t,
\end{align}
with $\velocities$ constant between events.

The event rate for the $i$-th coordinate is given by
\begin{align}
\lambda_i(\parameters, \velocities) = \max\left(0, \, \velocities_i \, \frac{\partial}{\partial \parameter_i} U(\parameters)\right),
\end{align}
where $U(\parameters) = -\log \prior{\parameters} - \log \likelihood{\parameters}$ is the negative log-posterior (up to an additive constant).
When an event occurs for coordinate $i$, the velocity of that coordinate flips sign: $\velocities_i \leftarrow -\velocities_i$, while all other velocities remain unchanged.
The overall event rate is the sum of the coordinate-wise rates: $\lambda(\parameters, \velocities) = \sum_{i=1}^d \lambda_i(\parameters, \velocities)$.

Intuitively, the Zig--Zag sampler moves each parameter at constant speed in one direction until the gradient of the log-posterior suggests that the density is decreasing.
At that point, the velocity flips, and the parameter reverses direction.
This produces a trajectory that ``zig-zags'' through the parameter space, and it can be shown that the resulting process has the target posterior as its stationary distribution \citep{bierkens2019ZigZag}.

\subsection{Bouncy Particle Sampler}

The Bouncy Particle sampler (BPS; \citealp{bouchard2018Bouncy}) generalizes the Zig--Zag sampler by allowing the velocity to take any value in $\reals^d$ rather than being restricted to $\{-1, +1\}^d$.
The velocity is typically initialized from a standard multivariate normal distribution and is refreshed from this distribution at random refreshment events.
Between events, the parameters evolve as
\begin{align}
\parameters(t + \Delta t) = \parameters(t) + \velocities \, \Delta t.
\end{align}

The event rate is given by
\begin{align}\label{eq:rate-bps}
\lambda(\parameters, \velocities) = \max\left(0, \, \velocities^\top \nabla U(\parameters)\right),
\end{align}
where $\nabla U(\parameters)$ is the gradient of the negative log-posterior.
When an event occurs, the velocity is reflected against the gradient:
\begin{align}
\velocities' = \velocities - 2 \, \frac{\velocities^\top \nabla U(\parameters)}{\|\nabla U(\parameters)\|^2} \, \nabla U(\parameters).
\end{align}
This reflection is analogous to a perfectly elastic collision with a hyperplane perpendicular to the gradient.
In addition to these gradient-based reflections, the Bouncy Particle sampler also includes refreshment events at a constant rate, where the velocity is resampled from its prior distribution.
The refreshment events help ensure ergodicity \citep{bouchard2018Bouncy}.

\subsection{Boomerang Sampler}

The Boomerang sampler \citep{bierkens2020boomerang} introduces a Gaussian reference distribution with mean $\parameters_\star$ and covariance $\Sigma_\star$ that encourages the process to revisit regions of high posterior density.
The parameters and velocities evolve according to
\begin{align}
\frac{d\parameters_t}{dt} = \velocities_t, \quad
\frac{d\velocities_t}{dt} = -(\parameters_t - \parameters_\star),
\end{align}
with solution $\parameters_t = \parameters_\star + (\parameters_0 - \parameters_\star)\cos(t) + \velocities_0\sin(t)$ and $\velocities_t = -(\parameters_0 - \parameters_\star)\sin(t) + \velocities_0\cos(t)$.

The Gaussian reference distribution is incorporated by expressing the posterior relative to this reference:
\begin{align}
U_{\mathrm{res}}(\parameters) =  U(\parameters) -
\frac{1}{2}(\parameters - \parameters_\star)^\top \Sigma_{\star}^{-1} (\parameters- \parameters_\star).
\end{align}
The posterior is then written as the product of the Gaussian reference density and $\exp(-U_{\mathrm{res}}(\parameters))$.
The event rate has the same form as for the Bouncy Particle sampler, but depends only on the residual $U_{\mathrm{res}}(\parameters)$:
\begin{align}
    \lambda(\parameters,\velocities)=\max(0,\velocities^\top\nabla U_{\mathrm{res}}).
\end{align}
The idea behind the Boomerang sampler is straightforward.
If the target posterior is close to the Gaussian reference distribution, then $U_{\mathrm{res}}(\parameters)$ varies little and reflection events are rare.
Instead the process is governed by refreshment events which resample the velocities from $\dnorm{\bm{0}, \Sigma_\star}$.
This is particularly convenient in combination with the Bernstein--von Mises Theorem \citep{vandervaart1998asymptotic} which implies that asymptotically many posterior distributions converge to a multivariate Gaussian as the sample size increases.

\begin{figure}[!ht]
    \includegraphics[width=\textwidth]{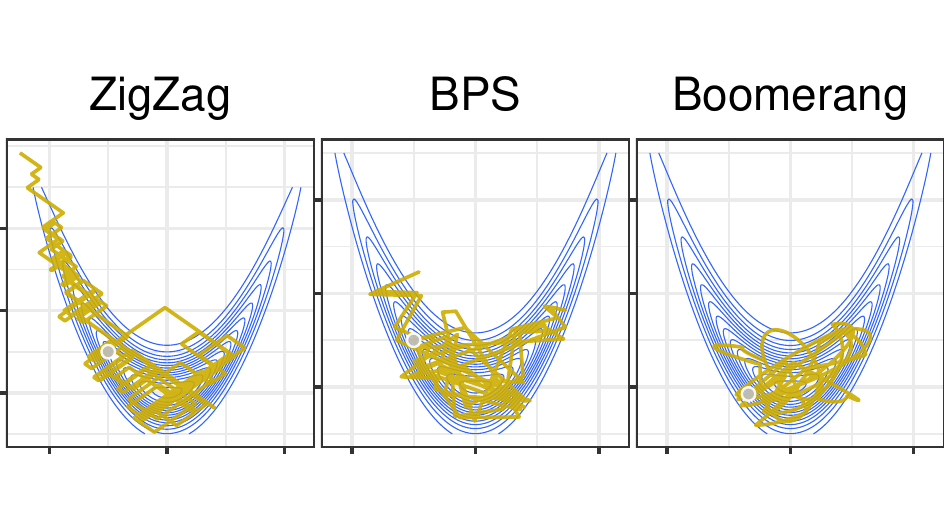}
    \caption{Illustration of the three PDMP dynamics considered. The simulation time is intentionally short to show the behavior of the three dynamics and not meant to illustrate convergence, for which a longer simulation would be needed.}\label{fig:illustration}
\end{figure}

Figure~\ref{fig:illustration} illustrates the behavior of the three dynamics on a short trajectory.
Evident is the jagged path of the Zig--Zag sampler (left panel) which can move in four directions, the Bouncy Particle sampler (middle panel), which can move in any direction but still moves linearly, and the Boomerang sampler (right panel) where the process moves in a smooth curved path due to the Gaussian reference.

\subsection{Simulating Event Times}

The key computational challenge for PDMPs is simulating the time until the next event.
There are three types of events that can occur: refreshments, reflections, and sticky events.
For each event type, we determine its candidate time, and the earliest event occurs.

Refreshments are typically the simplest events.
They occur at a constant predetermined rate and are obtained from a draw from an exponential distribution.
Reflection times instead follow an inhomogeneous Poisson process with rate $\lambda(\parameters(t+u),\velocities(t+u))$ along the deterministic trajectory.
The cumulative distribution function of the next event time $\Delta t$ is
\begin{align}\label{eq:event_times}
F(s) = 1 - \exp\left(-\int_0^s \lambda\big(\parameters(t+u), \velocities(t+u)\big) \, du\right).
\end{align}
Simulating from this distribution requires solving an integral equation that generally does not have a closed form.
Common approaches use Poisson thinning \citep{bierkens2019ZigZag}, which requires deriving an upper bound on the event rate.
We use a recent approach by \citet{andral2024automated} that evaluates the rate function and its derivative with respect to $t$ on a grid, thereby avoiding the need for explicit upper bounds.

The `sticking to zero' event is deterministic, and occurs when the deterministic trajectory of a parameter reaches zero.
For example, for the Zig--Zag and BPS, the hitting time for parameter $\parameter_i$ is $$t_i^{\text{hit}} = -\parameter_i / \velocity_i$$ if $\parameter_i\velocity_i < 0$, and $\infty$ otherwise.
A parameter stuck at zero leaves the spike after a random waiting time determined by the adjacent-model posterior mass.
For independent priors, this is an exponential waiting time with rate proportional to the prior inclusion odds and the slab density at zero.

\subsection{Sticky Events for Variable Selection}

The PDMP samplers described above are designed for continuous parameter spaces and cannot directly perform variable selection, which requires parameters to take the exact value zero with positive posterior probability.
\citet{bierkens2023sticky} introduced sticky events to extend PDMPs to spike-and-slab variable selection.

Each parameter $\parameter_i$ is associated with an inclusion indicator $\indicator_i \in \{0, 1\}$.
When $\indicator_i = 1$, the parameter evolves according to the standard PDMP dynamics.
When $\indicator_i = 0$, the parameter and its velocity are fixed at zero.
Thus, the process evolves on a mixed discrete-continuous state space: excluded parameters remain exactly at the spike, while included parameters move continuously under the chosen dynamics.

Sticky events govern transitions between these two states.
If $\indicator_i = 1$, the next possible exclusion event is deterministic: the parameter can only become excluded when the deterministic trajectory hits zero.
For linear dynamics, this hitting time is $t_i^{\text{hit}} = -\parameter_i / \velocity_i$ when $\parameter_i\velocity_i < 0$ and $\infty$ otherwise.
At this time the parameter reaches zero exactly.
If $\indicator_i = 0$, the parameter leaves the spike after a random waiting time.
For independent spike-and-slab priors, this is a Poisson process with fixed rate determined by the prior inclusion probability and the density of the slab distribution at zero.
Specifically, the rate is
\begin{align}\label{eq:unfreeze_indep}
    \kappa_i = C_v\rho_i f_i(0),
\end{align}
where $\rho_i = p(\indicator_i=1)/(1-p(\indicator_i=1))$ is the prior inclusion odds, $f_i(0)$ is the slab density at zero, and $C_v=\mathbb{E}|V_i|$ is the mean absolute departure speed which depends on the velocity distribution and thus the PDMP dynamics used.
For Zig--Zag, $C_v=1$, and for BPS and Boomerang, $C_v=\sqrt{2/\pi}$.

\subsection{Posterior Inference From PDMP Trajectories}
In discrete-time MCMC, posterior inference is typically based on the empirical distribution of the sampled states after a burn-in period.
Conveniently, any statistic of interest can be estimated by averaging its value over the sampled states, e.g., the posterior mean is just the mean of the raw posterior samples.
For PDMPs, the situation is more complicated because the process evolves continuously in time.
For example, for the Zig--Zag and BPS we have the following estimator:
\begin{align}
    \int g(\theta)\pi(\theta\mid\text{data}) \enspace\mathrm{d}\theta \approx \frac{1}{\tau_{N+1} - \tau_{0}}\sum_{i=0}^N\int_0^{\tau_{i+1} - \tau_{i}} g(x_{\tau_i} + s\velocities_{\tau_i}) \enspace\mathrm{d}s,
\end{align}
This integral depends on both the dynamics and the estimator of interest.
For common estimators, such as the posterior mean, and the three dynamics we consider here, this integral can be computed in closed form.
For more advanced quantities, we discretize the integral by evaluating the function at a grid of points along the deterministic trajectory.
We pick a number of grid points $M$ and define $h = (\tau_{N+1} - \tau_{0})/M$ and then use
\begin{align}
    \int g(\theta)\pi(\theta\mid\text{data}) \enspace\mathrm{d}\theta \approx \frac{1}{M}\sum_{i=1}^M g(x_{\tau_0 + ih}).
\end{align}
Increasing $M$ improves the accuracy of the approximation, but also increases the computational cost.

Some quantities of interest in discrete-time MCMC are less meaningful for PDMPs, in particular the autocorrelation.
For discrete-time MCMC, a high autocorrelation indicates that the chain is moving slowly through the parameter space and is often a sign of poor mixing \citep{geyer1992practical}.
For PDMPs, if we discretize the trace at a high frequency, we will see that the autocorrelation is very high because the process is moving continuously and thus the states at nearby time points are highly correlated.
At the same time, the process may be exploring the parameter space efficiently, so a high autocorrelation does not necessarily indicate poor mixing.
Instead, we examine the effective sample size (ESS) relative to computation time to assess the sampling efficiency.
We use the continuous time definitions for ESS as described in \citet{bierkens2023sticky}. % TODO: wasn't another paper by Bierkens?
Furthermore, we can examine the number of events per unit time in relation to the distance travelled by the process.
A high event rate relative to the distance travelled may indicate inefficient exploration of the parameter space \citep{bouchard2018Bouncy, sherlock2022discrete}.

\subsection{Dependent Priors}

Existing sticky PDMP methods assume that the spike-and-slab prior factorizes as a product of independent priors over parameters \citep{bierkens2023sticky}.
Independent spike-and-slab priors are restrictive because they require both the model prior and the slab distribution to factorize across parameters.
More general priors allow, for example, multiplicity adjustment through a prior on model size and dependence between included coefficients through a joint slab distribution.
A common approach is to decompose the joint prior as $p(\indicators,\parameters) = p(\indicators)p(\parameters\mid\indicators)$.
This allows a model prior, $p(\indicators)$, to control the inclusion probabilities and expected model size.
The slab prior, $p(\parameters\mid\indicators)$, controls how the included parameters depend on one another.
Extending sticky events to dependent priors is straightforward in principle.
The exclusion events are unchanged: we compute the time until a parameter reaches zero.
Let $A=\{i:\indicator_i=1\}$ be the active set.
For an excluded coefficient $i\notin A$, define the adjacent-model prior ratio $\rho_i(A)=p(A\cup\{i\})/p(A)$.
Let $f_i(0\mid\parameters_A,\psi)$ be the conditional slab density at zero given the active coefficients and slab hyperparameters $\psi$.
Then the unfreezing rate for any parameter is a function of the current state and active set and is given by
\begin{align}\label{eq:unfreeze_dep}
\kappa_i = C_v \rho_i(A) f_i(0\mid\parameters_A,\psi).
\end{align}
The two prior factors have distinct interpretations: $\rho_i(A)$ determines how much prior probability is assigned to entering the larger model, while $f_i(0\mid\parameters_A,\psi)$ determines how much continuous prior density in that model meets the spike at the boundary.

Critically, $\rho_i(A)$ now depends on the active set, while $f_i(0\mid\parameters_A,\psi)$ can additionally depend on the continuously evolving state.
Consequently, the unfreezing rate may vary with time along the deterministic trajectory.
In general, this requires simulating an inhomogeneous Poisson process, just like the reflection events.
We therefore focus on two often-used classes of dependent priors for which the resulting rates have useful structure.
The first class are size-dependent model priors, such as the beta-binomial prior \citep{scott2010bayes}. % TODO: find another example!
The second are Gaussian scale mixtures, which are commonly used slab priors.
This includes the multivariate Gaussian slab with a shared scale parameter, which is a special case of the multivariate Gaussian scale mixture.

\paragraph{Size-dependent model priors.}
For the Beta--Bernoulli model prior, the adjacent-model odds depend only on the current model size.
If the current model contains $k$ coefficients, the prior odds of adding an excluded coefficient are
\begin{align}
\label{eq:beta-bernoulli-addition-odds}
\frac{p(A\cup\{i\})}{p(A)}
=
\frac{a+k}{b+d-k-1}.
\end{align}
Because the active set, and hence $k$, is constant between sticky events, this model-prior contribution remains constant along each deterministic segment.
Thus, simulating the next unfreezing event is straightforward and remains a draw from an exponential distribution.

\paragraph{Gaussian scale-mixture slab priors.}
The slab priors are more complex.
Unlike the model prior, the conditional density at zero for a particular coefficient depends on the current active coefficients and their shared scale parameters.
Furthermore, this density changes as the deterministic dynamics evolve the active coefficients, and thus also depends on the dynamics used.
We focus on Gaussian scale mixtures, which are commonly used slab priors.
These are defined as
\begin{align}
\psi &\sim p(\psi), &
\parameters_A\mid\psi
&\sim \dnorm{\mu_A(\psi),\Sigma_{AA}(\psi)},
\qquad A\subseteq\{1,\ldots,d\},
\end{align}
where $\psi$ contains the mixing variables and $\mu_A(\psi)$ and $\Sigma_{AA}(\psi)$ are the subvectors and submatrices of a $d$-dimensional Gaussian distribution.
The mixing variables remain part of the PDMP state rather than being integrated out; an ordinary multivariate Gaussian slab is recovered as the special case in which $\psi$ is fixed.
For an excluded coefficient $i\notin A$, the ratio between the slab densities under the larger and current models can be written as
\begin{align}
\frac{
    p(\parameters_A,\parameter_i=0\mid\psi)
}{
    p(\parameters_A\mid\psi)
}
=
p(\parameter_i=0\mid\parameters_A,\psi).
\end{align}
Thus, the contribution of the slab prior to the unfreezing rate is simply the conditional density of the excluded coefficient at zero given the currently active coefficients.
For the Gaussian slab, this density follows directly from Gaussian conditioning.
Specifically,
\begin{align}
\parameter_i\mid\parameters_A,\psi
&\sim \dnorm{m_i,s_i^2}, \\
m_i
&=\mu_i+\Sigma_{iA}\Sigma_{AA}^{-1}
  (\parameters_A-\mu_A), &
s_i^2
&=\Sigma_{ii}-\Sigma_{iA}\Sigma_{AA}^{-1}\Sigma_{Ai},
\end{align}
where the dependence of the mean and covariance on $\psi$ is omitted.
Consequently, the conditional density at 0 is
\begin{align}
f_i(0\mid\parameters_A,\psi)
=\frac{1}{s_i}\phi\left(\frac{m_i}{s_i}\right),
\end{align}
with $\phi$ the standard normal density.

Although each excluded coefficient has its own time-varying unfreezing rate, it is unnecessary to simulate a separate clock for each coefficient.
The superposition property of Poisson processes allows us to simulate a single clock with rate equal to their sum.
Specifically, we sum their rates and simulate one aggregate unfreezing clock,
\begin{align}
\Lambda_A(t)
=C_v\sum_{i\notin A}\rho_i(A)
f_i\{0\mid\parameters_A(t),\psi(t)\}.
\end{align}
Thus, if $E\sim\mathrm{Exponential}(1)$, the next unfreezing time is obtained
from $\int_0^t\Lambda_A(u)\,du=E$.  At that event time, coefficient $i$ is
selected with probability proportional to
$\rho_i(A)f_i\{0\mid\parameters_A(t),\psi(t)\}$.
This is identical in distribution to simulating each coefficient's unfreezing time separately and selecting the first to occur.

For some combinations of Gaussian slab and deterministic dynamics, the integrated hazard can be evaluated analytically or reduced to a simple numerical problem.
Otherwise, we evaluate the cumulative hazard numerically and invert it to obtain the next event time.
% The corresponding expressions for the dynamics considered here are given in Appendix~\ref{app:dependent-parameter-priors}.

\subsection{Subsampled Gradients}

The computational cost of a PDMP sampler can be reduced by replacing the full likelihood gradient with a subsampled estimator, while retaining the correct target distribution when the corresponding event times are simulated appropriately \citep{bierkens2019ZigZag, pakman2017stochastic}.
Suppose that the negative log-posterior can be written as
\begin{align}\label{eq:subsample}
U(\parameters)
=
U_{\mathrm{prior}}(\parameters)
+
\sum_{a=1}^L U_a(\parameters),
\end{align}
and define $g_a(\parameters)=\nabla U_a(\parameters)$.
If factors $J_1,\ldots,J_m$ are sampled independently with probabilities $\pi_a$, then
\begin{align}
\label{eq:subsampled-gradient}
\widehat{\nabla U}(\parameters)
=
\nabla U_{\mathrm{prior}}(\parameters)
+
\frac{1}{m}
\sum_{b=1}^{m}
\frac{g_{J_b}(\parameters)}{\pi_{J_b}}
\end{align}
is an unbiased estimator of the full gradient.
The prior contribution is evaluated exactly and is therefore neither subsampled nor rescaled.

Importantly, unbiasedness of the gradient estimator alone is not sufficient.
Because the PDMP event rate is a nonlinear function of the gradient, the subsampled estimator must be combined with an event-time construction that preserves the correct invariant distribution.
The precise construction depends on the PDMP dynamics and the factorization of the likelihood.

For the two models we consider here, subsampling is particularly useful for the OMRF where the observations are assumed to be independent and identically distributed.
We derive the corresponding subsampling construction in Appendix~\ref{app:omrf-clocks}.

\subsection{Implementation}
We provide the methods through the \texttt{PDMPSamplersR} R package and the \texttt{PDMPSamplers.jl} Julia package.
The R package prepares the model and data and returns posterior summaries.
The Julia package implements the PDMP dynamics, event-time simulation, adaptation, and subsampling.
In the analyses, the statistical models are specified in Stan \citep{Stan} and evaluated using BridgeStan \citep{RoualdesEtAl_2023_BridgeStan} which provides an interface to evaluate likelihood and gradient of Stan models.

\section{Models}

We evaluate the proposed PDMP variable selection methods on two psychometric models: the random intercept cross-lagged panel model (RI-CLPM; \citealp{hamaker2015critique}) and the ordinal Markov random field (OMRF; \citealp{MarsmanEtAl_2023_ordinal}).
These models represent two common types of psychometric data: longitudinal panel data and cross-sectional network data.
Both models feature a potentially large number of parameters and a natural sparsity structure, making them well-suited for variable selection.

\subsection{Random Intercept Cross-Lagged Panel Model}

The random intercept cross-lagged panel model (RI-CLPM; \citealp{hamaker2015critique}) is a longitudinal model that decomposes repeated measures into stable between-person differences and within-person dynamics.
For $N$ individuals measured on $p$ variables at $T$ time points, the model is specified as
\begin{align}
\bm{y}_{it} &= \bm{\mu}_t + \mathbf{b}_i + \mathbf{z}_{it},
\end{align}
where $\bm{y}_{it} \in \reals^p$ is the vector of observations for individual $i$ at time $t$, $\bm{\mu}_t \in \reals^p$ is a time-specific mean vector, $\mathbf{b}_i \sim \dnorm{0, \Sigma_b}$ captures stable between-person differences, and $\mathbf{z}_{it}$ captures within-person fluctuations.

The within-person component follows a first-order vector autoregressive process:
\begin{align}
\mathbf{z}_{it} = \mathbf{B} \, \mathbf{z}_{i, t-1} + \boldsymbol{\varepsilon}_{it}, \quad \boldsymbol{\varepsilon}_{it} \sim \dnorm{0, \Sigma_\varepsilon},
\end{align}
where $\mathbf{B} \in \reals^{p \times p}$ is the matrix of cross-lagged and autoregressive effects.
The diagonal elements of $\mathbf{B}$ represent autoregressive effects (the influence of a variable on itself at the next time point), while the off-diagonal elements represent cross-lagged effects (the influence of one variable on another at the next time point).

Variable selection in the RI-CLPM focuses on the cross-lagged effects in $\mathbf{B}$.
With $p$ variables there are $p(p-1)$ possible cross-lagged effects, many of which may be absent.
We therefore place a spike-and-slab prior on each off-diagonal element of $\mathbf{B}$.

We consider two prior specifications.
The first assigns independent priors to the cross-lagged effects, given by
\begin{align}
\indicator_{jk} &\sim \text{Bernoulli}(\rho), \quad j \neq k, \\
\predictor_{jk} \mid \indicator_{jk} = 1 &\sim \dnorm{0, \tau^2}, \\
\predictor_{jk} \mid \indicator_{jk} = 0 &= 0,
\end{align}
where $\rho$ is the prior inclusion probability and $\tau^2$ is the slab variance.
The second prior introduces dependence both among the inclusion indicators and among the included effects.
For the indicators, we use a Beta--Bernoulli prior:
\begin{align}
    \rho &\sim \mathrm{Beta}(\alpha_0,\beta_0), &
    \indicator_{jk}\mid\rho &\sim \mathrm{Bernoulli}(\rho).
\end{align}
After integrating out $\rho$, the inclusion indicators are dependent.
The resulting prior allows uncertainty about the overall sparsity level \citep{ScottBerger_2010}.
To additionally introduce dependence between the magnitudes of the included cross-lagged effects, we use an exchangeable Gaussian slab.
Writing $d=p(p-1)$ for the number of selectable cross-lagged effects and $\boldsymbol{\beta}_A$ for the active effects, the dependent slab is the
restriction of a $d$-dimensional exchangeable Gaussian distribution,
\begin{align}
\boldsymbol{\beta} \sim \dnorm{\mathbf{0},\,\tau^2\{(1-r)I_d+r\mathbf{1}_d\mathbf{1}_d^\top\}},
\end{align}
where $0\leq r<1$.
Restricting this full-dimensional distribution to each active set ensures that the slab distributions for adjacent models are mutually compatible.
It also gives the conditional boundary densities required by the sticky sampler.

After marginalizing the person-specific random intercepts, each individual's stacked \(pT\)-dimensional response vector follows a multivariate Gaussian distribution.
Consequently, the likelihood and its gradient can be evaluated using only the sample count, response sum, and response cross-product matrix.

\subsection{Ordinal Markov Random Field}

The ordinal Markov random field (OMRF; \citealp{MarsmanEtAl_2023_ordinal}) is a network model for ordinal data.
For $p$ observed ordinal variables, each taking values in $\{0, 1, \ldots, K-1\}$, the OMRF models the joint distribution using a Markov random field with pairwise interactions:
\begin{align}
p(\mathbf{x} \mid \boldsymbol{\mu}, \boldsymbol{\Sigma}) = \frac{1}{Z(\boldsymbol{\mu}, \boldsymbol{\Sigma})} \exp\left(\sum_{j=1}^p \sum_{r=0}^{K-1}\mathbb{I}(x_j=r) \mu_{jr} + \sum_{j=1}^p \sum_{k=j+1}^p x_j x_k \sigma_{jk}\right),
\end{align}
where $\mathbf{x} \in \{0, 1, \ldots, K-1\}^p$ is a vector of ordinal observations, $\boldsymbol{\mu}=\{\mu_{jr}:j=1,\ldots,p;\ r=1,\ldots,K-1\}$ contains the category-specific main effects, and $\boldsymbol{\Sigma}=(\sigma_{jk})\in\reals^{p\times p}$ is symmetric with a zero diagonal.
The normalizing constant $Z(\boldsymbol{\mu}, \boldsymbol{\Sigma})$ is intractable.
We therefore follow \citet{MarsmanEtAl_2023_ordinal} and replace the full likelihood by the product of conditional distributions:
\begin{align}
\likelihood{\mathbf{x} \mid \boldsymbol{\mu}, \boldsymbol{\Sigma}} \approx \prod_{j=1}^p p(x_j \mid \mathbf{x}_{-j}, \boldsymbol{\mu}, \boldsymbol{\Sigma}).
\end{align}
For person $n$ and node $j$, the conditional distribution is
\begin{align}
p(x_{nj}=r\mid\mathbf{x}_{n,-j})
=
\frac{\exp\{\mu_{jr}+r\sum_{k\neq j}x_{nk}\sigma_{jk}\}}
{\sum_{u=0}^{K-1}\exp\{\mu_{ju}+u\sum_{k\neq j}x_{nk}\sigma_{jk}\}},
\qquad r=0,\ldots,K-1,
\end{align}
where $\mu_{j0}$ is fixed to $0$ to identify the model.

The interaction parameters $\sigma_{jk}$ determine the conditional dependence structure of the network.
If $\sigma_{jk}=0$, variables $j$ and $k$ are conditionally independent given all remaining variables; if $\sigma_{jk}\neq0$, an edge is present between them.

Variable selection in the OMRF involves determining which edges $\sigma_{jk}$ are nonzero.
The number of pairwise interactions is $p(p-1)/2$ and thus grows quadratically in the number of variables.
In the original implementation by \citet{MarsmanEtAl_2023_ordinal}, an independent spike-and-slab prior was used with $\indicator_{jk} \sim \text{Bernoulli}(\rho)$ and $\sigma_{jk} \mid \indicator_{jk} = 1 \sim \mathrm{Cauchy}(0,1)$.

Here we consider a dependent alternative.
We use the same Beta--Bernoulli model prior as in the RI-CLPM, which induces dependence among the inclusion indicators.
We additionally introduce node-specific scale parameters that are shared among edges incident on the same node.
This allows nodes to differ in the typical magnitude of their incident edge weights, without imposing that the signs of these edges agree.
Specifically,
\begin{align}
\tau &\sim \mathrm{HalfNormal}(0,s_\tau^2), \\
\lambda_j &\sim \mathrm{HalfNormal}(0,s_\lambda^2), \quad j=1,\ldots,p, \\
\sigma_{jk}\mid\indicator_{jk}=1,\tau,\lambda_j,\lambda_k
&\sim \dnorm{0,\,s_0^2\tau^2\lambda_j\lambda_k}.
\end{align}
Conditional on the scale parameters, the included edges are independent Gaussian variables.
Marginally, however, edges incident on the same node are dependent because they share the corresponding node-specific scale.

This construction is related to global--local shrinkage priors, such as the horseshoe \citep{carvalho2010horseshoe} and graphical horseshoe \citep{li2019graphical}, which also express coefficients as Gaussian scale mixtures.
Here the local scales are associated with nodes rather than individual edges, thereby sharing shrinkage information across incident edges.

This formulation is particularly convenient for the dependent sticky PDMP developed above.
Conditional on the scale parameters, the slab density of an excluded edge at zero is
\begin{align}
f_{jk}(0\mid\tau,\lambda_j,\lambda_k) = \frac{1}{\sqrt{2\pi}\,s_0\tau\sqrt{\lambda_j\lambda_k}}.
\end{align}
Consequently, the unfreezing rate follows directly from Equation~\ref{eq:unfreeze_dep}, while changes in the node-specific and global scales are automatically reflected in the time-varying boundary rate.

\section{Simulation Study}

We evaluate the proposed sticky PDMP methods in two simulation studies, using the RI-CLPM and OMRF as complementary test cases.
For each model, we compare the Zig--Zag, Bouncy Particle, and Boomerang dynamics, with reversible-jump MCMC implemented in NIMBLE serving as a condition-matched reference where feasible \citep{nimble-article:2017}.

The simulation study addresses four questions.
First, how efficiently does each method explore the posterior, as measured by the number of effective samples produced per second?
Second, how accurately do the methods recover the underlying sparse structure?
Third, how accurately do posterior means and credible intervals recover the true parameter values?
Fourth, when different samplers target the same posterior distribution, to what extent do their posterior summaries agree?

The two model families emphasize different computational aspects of the proposed methodology.
For the RI-CLPM, the Gaussian likelihood can be reduced to fixed-dimensional sufficient statistics after marginalizing the person-specific effects.
This setting therefore allows us to focus on the influence of the PDMP dynamics and the use of independent versus dependent spike-and-slab priors without requiring stochastic gradients.
For the OMRF, in contrast, the pseudolikelihood decomposes into person--node contributions.
This makes it possible to study whether unbiased gradient subsampling improves computational efficiency, in addition to comparing independent and dependent prior specifications.

\subsection{Design}

For the RI-CLPM, we fix $p=4$ variables measured at $T=4$ time points and vary the sample size between $N=100$ and $N=300$.
The autoregressive effects are fixed at $0.35$.
The 12 selectable cross-lagged effects follow one of three structures: a null pattern, a common-sender pattern, or a mixed-sign pattern.
Nonzero cross-lagged effects have absolute magnitude $0.15$.
Each simulation condition was replicated 10 times.

We compare two prior specifications.
The independent prior uses a fixed inclusion probability of $0.20$ and a Gaussian slab with standard deviation $0.35$.
The dependent prior combines a Beta$(1,4)$ inclusion hierarchy with an exchangeable Gaussian slab with marginal standard deviation $0.35$ and correlation $0.25$.
Under both prior specifications, we compare the Zig--Zag, Bouncy Particle, and Boomerang dynamics with the condition-matched NIMBLE reference.

For the OMRF, each variable has $K=4$ response categories.
We vary the sample size between $N=200$ and $N=500$ and the number of nodes between $p=10$, $20$, and $30$.
The generated networks have edge density $0.20$ and follow either a random or hub-structured topology.
Nonzero edges have absolute magnitude $0.25$ and random signs.
Each simulation condition was replicated 10 times.

The OMRF simulation compares full-gradient evaluation with person--node subsampling for each of the three PDMP dynamics under both prior specifications.
The independent specification combines a Beta$(1,4)$ model prior with a standard Cauchy slab.
The dependent specification combines the same model prior with the node-shared Gaussian scale-mixture slab.
A condition-matched NIMBLE reference is included for both priors when $p=10$.
The larger OMRF conditions are used for the efficiency and structure-recovery comparisons but not for posterior-agreement comparisons with NIMBLE.
All OMRF PDMP runs use 50 time units of warmup.
The retained trajectory is 500 time units for Zig--Zag and BPS and 50 time units for Adaptive Boomerang.

% Reported sampler time excludes compilation time.
% The PDMP samplers use diagonal preconditioning that is adapted during warmup and then held fixed during the retained trajectory.
% Initialization, warmup, retained sampling, gradient evaluation, and total wall-clock time are recorded separately.
% The NIMBLE reference uses its own warmup and sampling settings and is used primarily to assess posterior agreement rather than as a software-independent benchmark of computational efficiency.
We evaluate the samplers in terms of computational efficiency, structure recovery, parameter estimation, and posterior agreement.
Our primary efficiency measure is the median effective sample size of the selectable parameters per second of retained sampling time.
For the PDMP samplers, the continuous-time trajectory is weighted by residence time when computing the effective sample size.
We additionally record wall-clock time and the number of gradient evaluations.
For subsampled runs, the number of gradient evaluations is adjusted by the fraction of observations used.

Structure recovery is evaluated using posterior inclusion probabilities.
For conditions containing both present and absent effects, we report the area under the receiver operating characteristic curve.
At a posterior inclusion threshold of $0.50$, we additionally report sensitivity, specificity, precision, false-positive rate, and balanced accuracy.
Parameter estimation is evaluated using posterior means, root mean squared error, and credible-interval coverage.
Finally, we assess posterior agreement by comparing posterior inclusion probabilities, posterior means, and posterior model size.

% \subsection{Performance Measures}

% The primary efficiency measure is the median selectable-parameter effective sample size per main sampler second.
% The continuous-time trajectory is weighted by residence time before effective sample size is computed.
% We also report wall-clock time, gradient calls, subsampled factor evaluations, and full-data anchor evaluations.

% The primary selection measure is the area under the receiver operating characteristic curve based on posterior inclusion probabilities.
% Area under the curve is undefined for the all-zero condition and is not interpreted there.
% At a posterior inclusion threshold of 0.50, we report sensitivity, specificity, false-positive rate, precision, and balanced accuracy.

% Posterior agreement is assessed only against a condition-matched reference with the same likelihood or pseudolikelihood and the same prior.
% We compare posterior inclusion probabilities, posterior means, and posterior model size.
% Selection accuracy and posterior agreement answer different questions and are reported separately.

\subsection{Results}

\subsubsection{RI-CLPM}

\begin{figure}[!ht]
\centering
\includegraphics[width=0.92\textwidth]{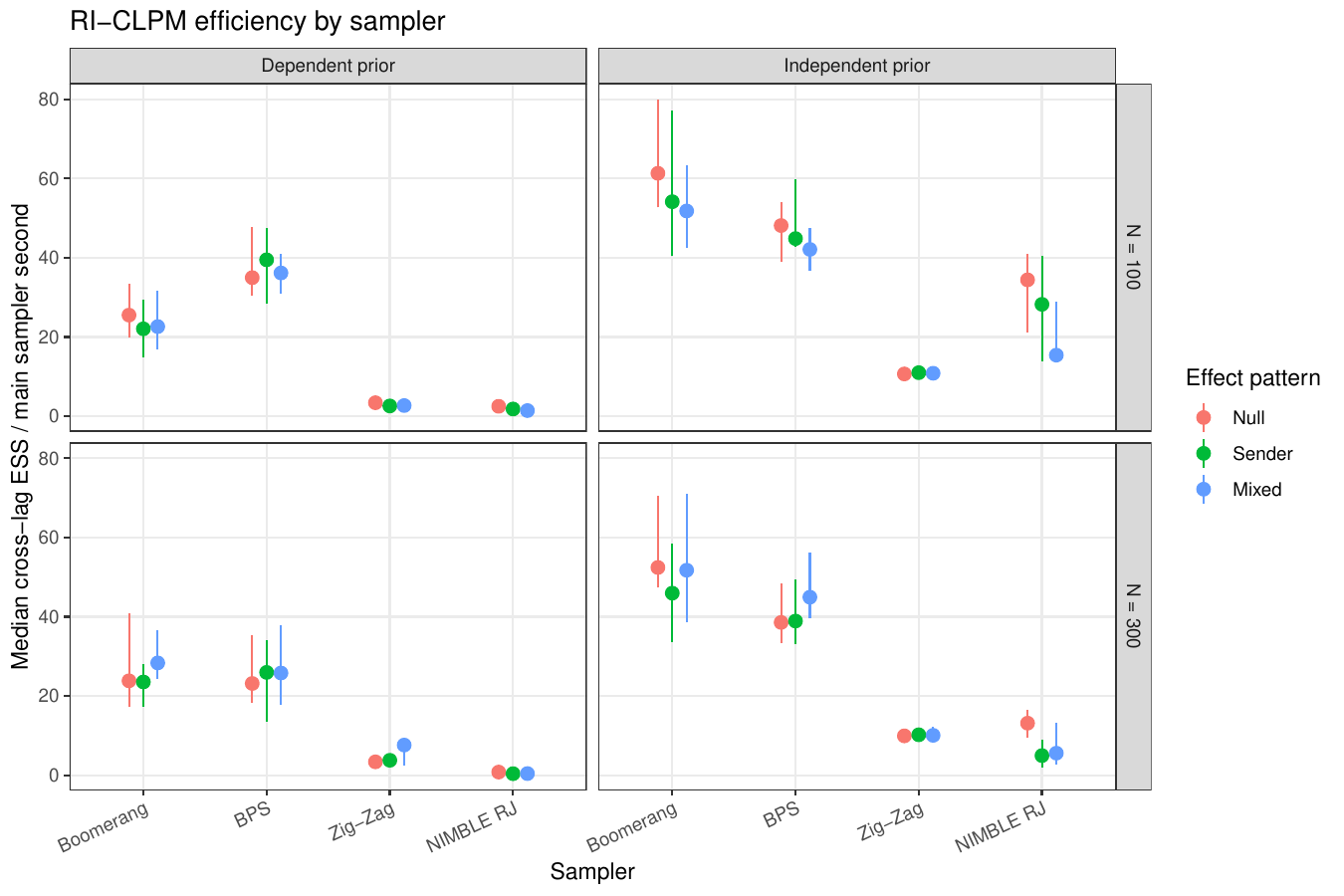}
\par\smallskip
\caption{RI-CLPM computational efficiency.
Each point is the median over 10 replications for one sample size, prior, effect pattern, and sampler.
}
\label{fig:clpm-efficiency}
\end{figure}

All 480 planned RI-CLPM fits completed, corresponding to 10 replications for every combination of sample size, effect pattern, prior, and sampler.
No fits were excluded.
The BPS and Boomerang were the most efficient samplers.
Across conditions, their median effective sample sizes per second of retained sampling time were 38.1 and 35.7, respectively, compared with 8.66 for the Zig--Zag and 3.02 for NIMBLE.
The relative performance of BPS and Boomerang depended somewhat on sample size: BPS was more efficient at $N=100$, whereas Boomerang was slightly more efficient at $N=300$.
The efficiency of all three PDMP samplers changed little as the sample size increased, whereas the median ESS rate of NIMBLE decreased from 7.89 at $N=100$ to 1.58 at $N=300$.
This is consistent with the use of fixed-dimensional sufficient statistics for evaluating the RI-CLPM likelihood and gradient.
Figure~\ref{fig:clpm-efficiency} shows the efficiency for each condition aggregated over repetitions.

Structure recovery improved substantially with sample size.
Across samplers and non-null conditions, the median area under the receiver operating characteristic curve increased from 0.85 at $N=100$ to 1.00 at $N=300$.
At a posterior inclusion threshold of $0.50$, median sensitivity increased from 0.33 to 1.00, while median specificity was 1.00 at both sample sizes.
In the null conditions, only 9 of the 1,920 decisions concerning truly absent cross-lagged effects resulted in false positives.
Parameter estimation showed a similar improvement, with median cross-lagged RMSE decreasing from approximately 0.049 at $N=100$ to 0.022 at $N=300$ across all four samplers.

\begin{figure}[!ht]
\centering
\includegraphics[width=0.92\textwidth]{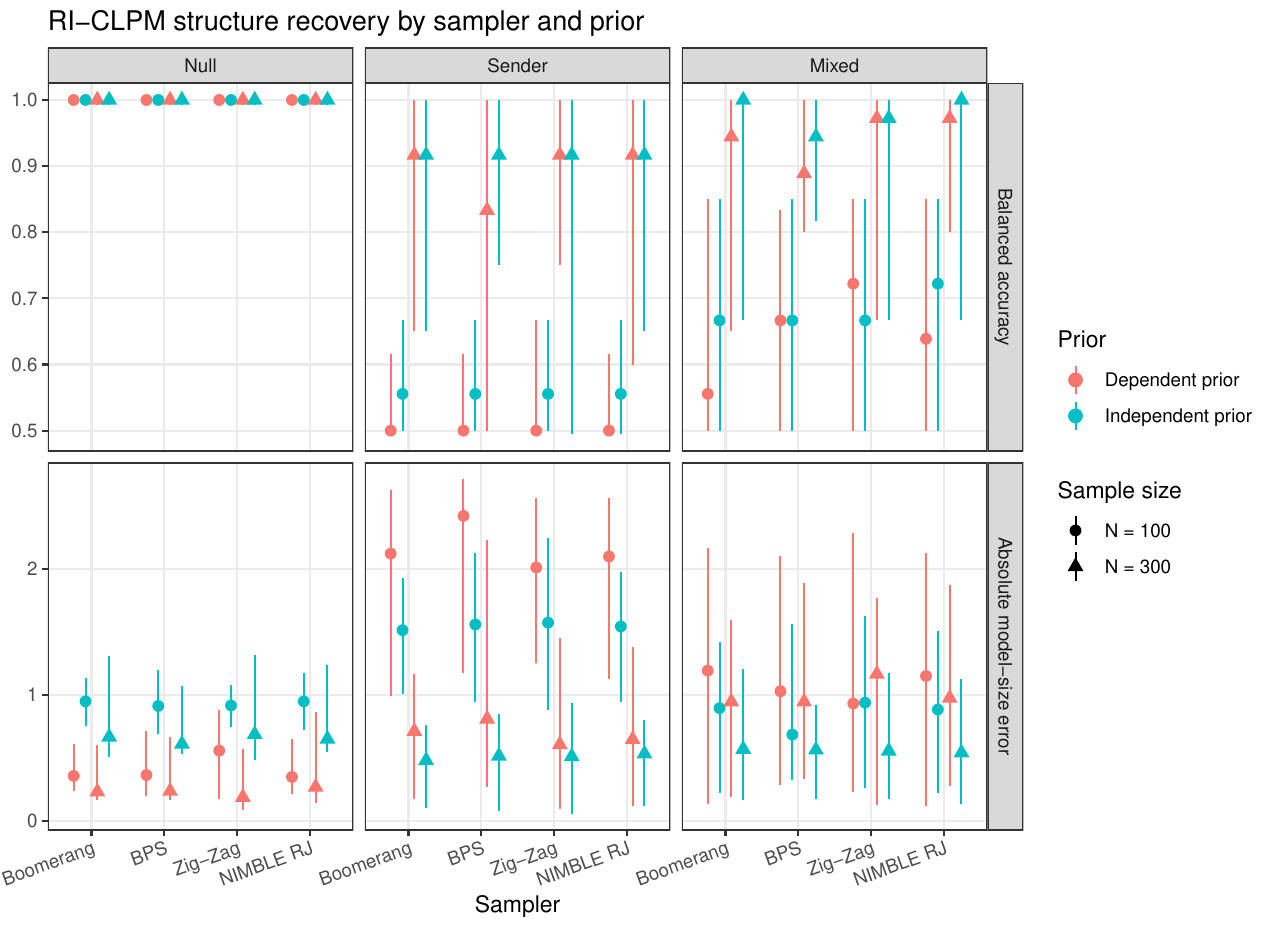}
\par\smallskip
\caption{RI-CLPM structure recovery by prior.
Balanced accuracy and absolute posterior model-size error use a posterior inclusion threshold of $0.50$.
For the null condition, balanced accuracy reduces to specificity because sensitivity is undefined.
The horizontal axis separates Adaptive Boomerang, BPS, Zig--Zag, and NIMBLE.
Colour denotes the prior, and point shape denotes sample size.
}
\label{fig:clpm-structure-recovery}
\end{figure}

The dependent prior primarily affected posterior sparsity.
In the null condition, it reduced the median posterior model size from 0.93 to 0.38 at $N=100$ and from 0.64 to 0.22 at $N=300$.
This improvement did not extend uniformly to the non-null conditions.
At $N=100$, the dependent prior reduced model-size error under the null but also reduced balanced accuracy in the common-sender and mixed-sign conditions.
These differences became smaller at $N=300$.
Figure~\ref{fig:clpm-structure-recovery} illustrates this trade-off.

Finally, the three PDMP dynamics agreed closely with the condition-matched NIMBLE reference.
The median mean absolute differences in posterior inclusion probabilities were 0.019 for Adaptive Boomerang, 0.028 for BPS, and 0.028 for Zig--Zag.
The corresponding median mean absolute differences in posterior means were 0.0023, 0.0033, and 0.0030.
These results indicate that the PDMP samplers target the same posterior distribution as the reversible-jump reference while providing substantially higher sampling efficiency in this setting.

\subsubsection{OMRF}

\begin{figure}[!ht]
\centering
\includegraphics[width=0.92\textwidth]{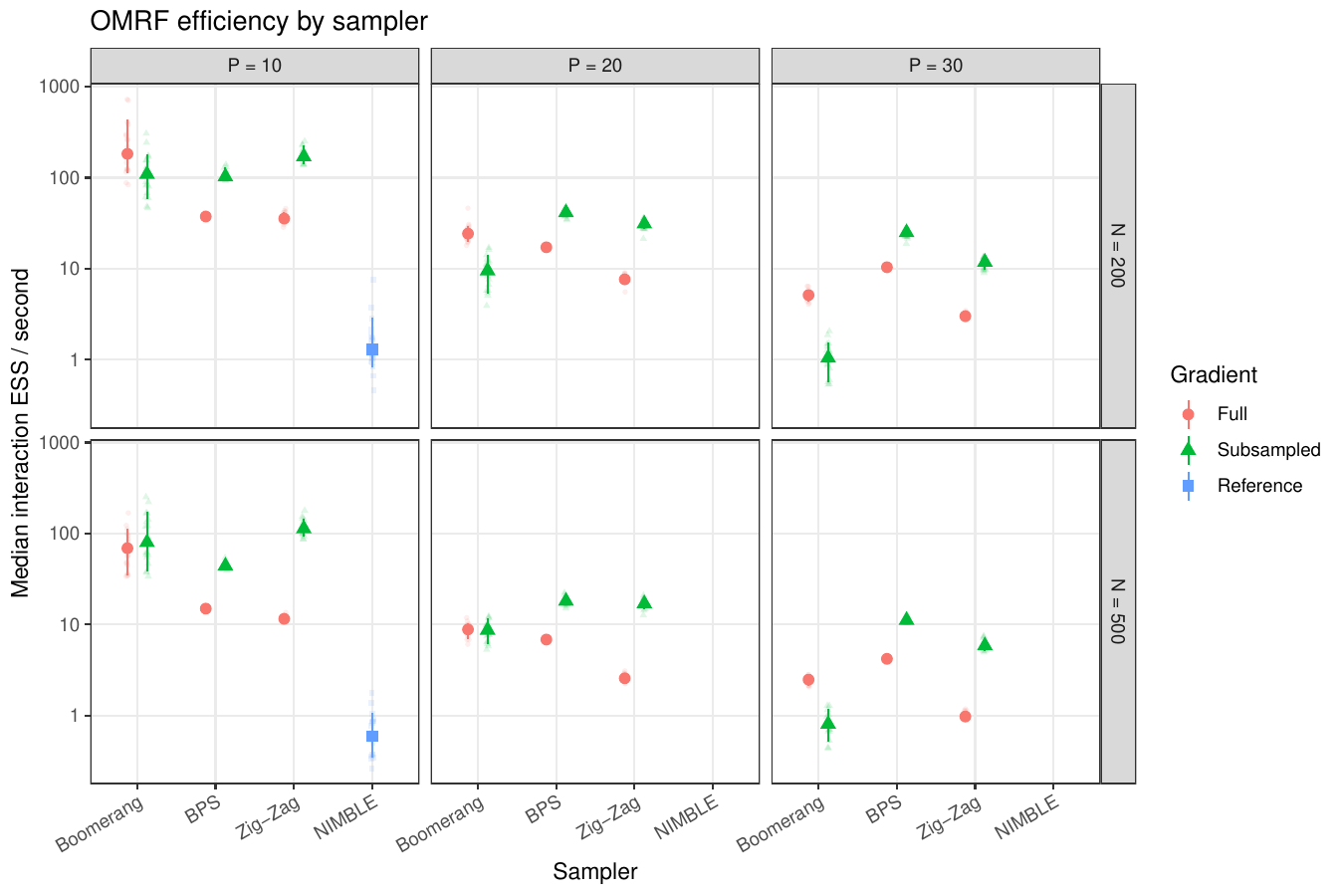}
\par\smallskip
\caption{OMRF computational efficiency under the independent-slab gradient comparison.
Small points show individual replications.
The y-axis is logarithmic.}
\label{fig:omrf-efficiency}
\end{figure}

All 1,520 planned OMRF fits completed, consisting of 1,440 PDMP fits and 80 NIMBLE reference fits.
No fits were excluded.
Computational efficiency decreased as the number of nodes increased.
The effect of subsampling depended strongly on the PDMP dynamics.
For BPS, the median subsampled-to-full ESS-rate ratios were 2.90, 2.52, and 2.34 for $p=10$, $20$, and $30$, respectively.
For Zig--Zag, the corresponding ratios were 4.35, 2.93, and 2.10.
Subsampling did not improve Boomerang, with ratios of 1.00, 0.95, and 0.54.
Figure~\ref{fig:omrf-efficiency} shows the efficiency results across conditions.

Although the subsampled Boomerang produced more raw effective samples than the full-gradient Boomerang in some conditions, this did not compensate for the additional computational cost.
At $p=30$ and $N=500$, for example, the median number of effective samples increased from 106 to 158, while median sampling time increased from 42 to 208 seconds.
The lack of an efficiency gain therefore appears to be specific to the current Boomerang event-time implementation rather than to subsampling itself.

\begin{figure}[!ht]
\centering
\includegraphics[width=0.92\textwidth]{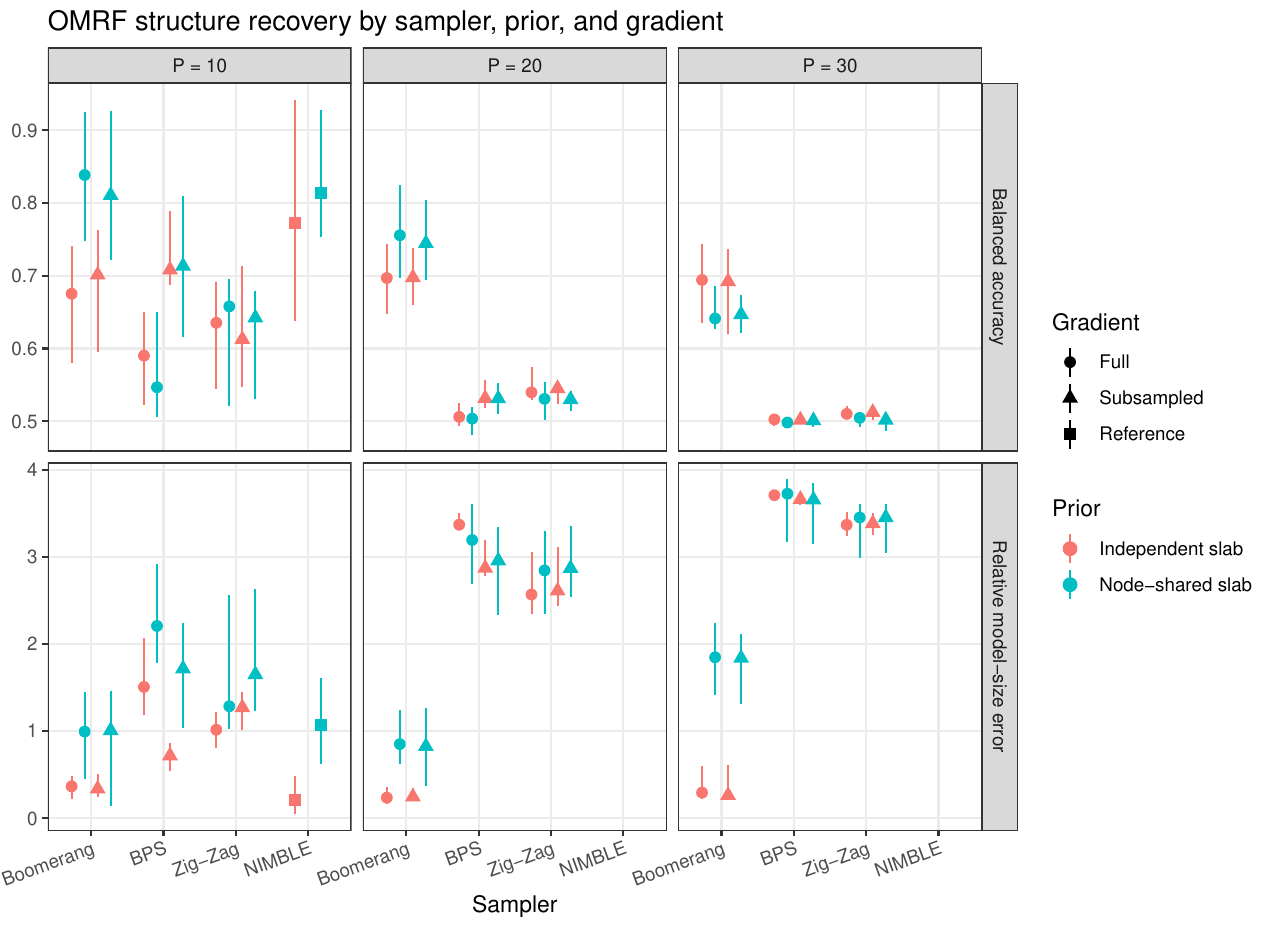}
\par\smallskip
\caption{OMRF structure recovery by prior.
Balanced accuracy uses a posterior inclusion threshold of $0.50$.
Relative model-size error is the absolute difference between posterior expected and true model size, divided by true model size.
The horizontal axis separates Boomerang, BPS, Zig--Zag, and NIMBLE.
Colour denotes the prior, and point shape denotes the full-gradient, subsampled, or NIMBLE reference implementation.
Points are medians over 10 replications, with sample sizes and graph structures averaged within replication.
NIMBLE was fitted only at $p=10$.}
\label{fig:omrf-structure-recovery}
\end{figure}

Structure recovery became more difficult as the number of nodes increased.
At $p=30$ under the independent slab, median balanced accuracy was 0.69 for Boomerang and approximately 0.50 for BPS and Zig--Zag.
The corresponding relative model-size errors were approximately 0.3, 3.7, and 3.4.
Within each dynamics, full and subsampled gradients produced similar structure-recovery results.
The differences therefore appear to be driven mainly by the PDMP dynamics rather than by gradient subsampling.
Figure~\ref{fig:omrf-structure-recovery} shows the corresponding comparisons.

Posterior agreement was weaker than in the RI-CLPM study.
The median mean absolute differences in posterior inclusion probabilities between full and subsampled gradients were 0.108 for Zig--Zag, 0.167 for BPS, and 0.120 for Boomerang.
For the full-gradient fits at $p=10$, the corresponding differences from NIMBLE were 0.351, 0.424, and 0.140.
These differences were not fully explained by the recorded Monte Carlo uncertainty.
The OMRF results should therefore be interpreted with some caution until the source of this disagreement is resolved.

\section{Empirical Example}

As an empirical illustration, we fitted the OMRF to the 14 Mental Health Continuum--Short Form items in the first wave of the LISS panel \citep{lamers2011evaluating, ScherpenzeelDas2010}.
The analysis was based on complete responses from 1,804 participants.
Each item had six ordered response categories.
We used the dependent prior together with the full-gradient Adaptive Boomerang sampler.
The first 500 time units were discarded as warmup, and the subsequent 2,000 time units were retained for posterior inference.

The fitted model favored a dense network structure.
All 91 edge inclusion probabilities exceeded .50, and the posterior expected model size was 85.3 edges.
Moreover, 65 edges had posterior inclusion probabilities of at least .95, and for 60 edges the 95\% credible interval excluded zero.
The strongest posterior mean associations were between happiness and satisfaction (.75), personality satisfaction and daily responsibility (.37), interest in life and satisfaction (.34), and happiness and interest in life (.33).
Monte Carlo precision was adequate, with a minimum effective sample size of 796 across edges and a median of 1,878.

\begin{figure}[!ht]
\centering
\includegraphics[width=.82\textwidth]{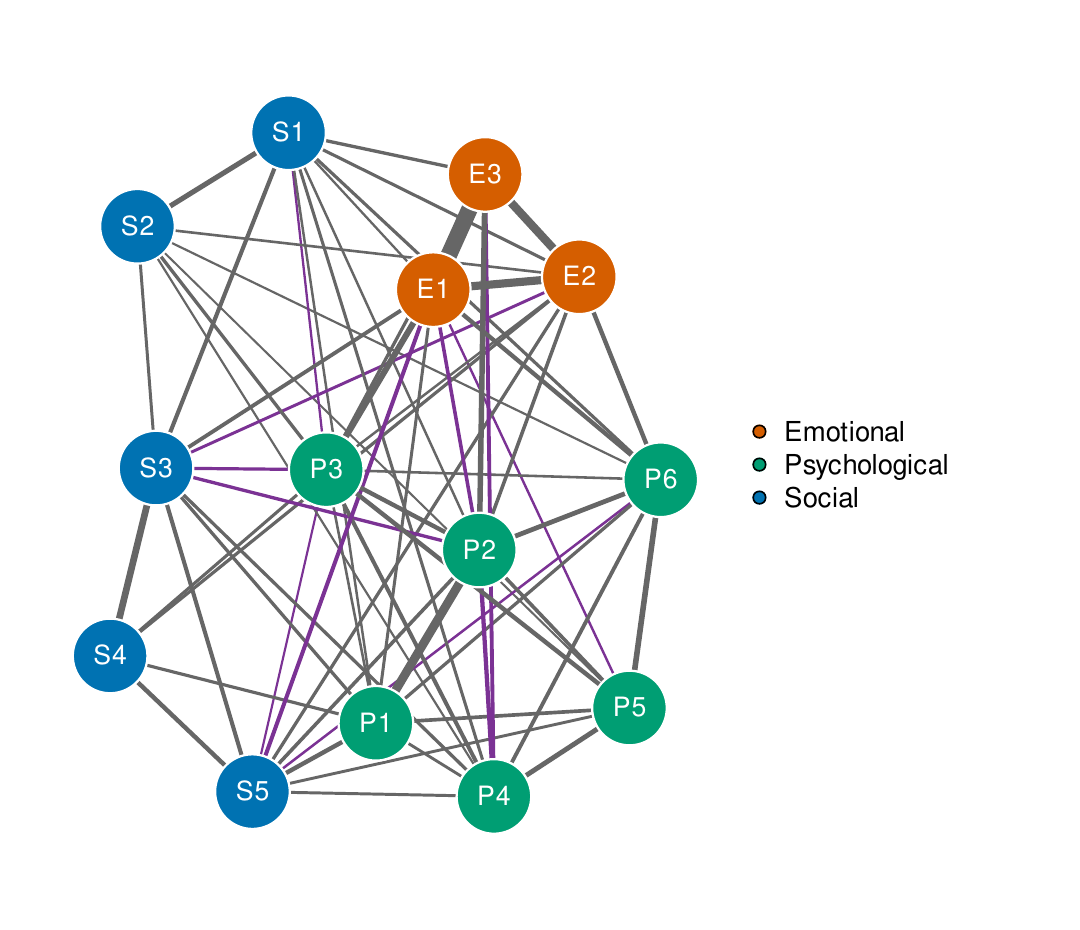}
\caption{Posterior mean OMRF network from the Adaptive Boomerang fit.
Edges are shown when their posterior inclusion probability is at least .95.
Edge width represents the absolute posterior mean, and edge color indicates its sign.
The labels E1--E3, S1--S5, and P1--P6 denote emotional, social, and psychological well-being items.}
\label{fig:empirical-omrf-network}
\end{figure}

Figure~\ref{fig:empirical-omrf-network} displays the high-probability part of the estimated network.
Overall, the empirical example suggests that for these well-being items the posterior favors a dense weighted association structure rather than a sparse network with only a small number of selected edges.

\section{Discussion}

We developed a sticky PDMP framework for Bayesian variable selection with dependent priors and applied it to two psychometric models with different computational structures.
The first contribution is an extension of existing sticky PDMP methods beyond independent spike-and-slab priors \citep{bierkens2023sticky}.
By expressing the unfreezing rate in terms of an adjacent-model prior ratio and a conditional slab density at zero, the framework accommodates dependence both among inclusion indicators and among the included parameters.
This includes model-size priors such as the Beta--Bernoulli prior and Gaussian scale-mixture slabs.
The second contribution concerns gradient evaluation.
When the likelihood decomposes into many factors, as in the OMRF pseudolikelihood, unbiased subsampling can be incorporated while preserving the stated target distribution, provided that the corresponding event-time construction is valid \citep{fearnhead2018piecewise, bierkens2019ZigZag}.

The simulation results show that these constructions can provide an efficient alternative to conventional reversible-jump sampling.
For the RI-CLPM, the PDMP samplers produced substantially more effective samples per second than the NIMBLE reference, while giving closely matching posterior inclusion probabilities and posterior means.
Their efficiency also changed little as the sample size increased, consistent with the fixed-dimensional sufficient-statistic representation of the likelihood.
For the OMRF, subsampling substantially improved the efficiency of some of the PDMP implementations as the number of nodes increased.
At the same time, the weaker posterior agreement observed in the OMRF study makes clear that computational efficiency should not be considered independently of target agreement and Monte Carlo accuracy.
The empirical OMRF example further illustrates that the method is not restricted to settings in which the posterior strongly favors sparse models: in the well-being data, the posterior instead supported a relatively dense network.

Several limitations remain.
First, the dependent-prior extension is not completely general.
The sticky boundary rate requires the prior under adjacent models to be related in a way that permits evaluation of the corresponding model-prior ratio and conditional slab density.
The Gaussian constructions considered here make these quantities tractable, but arbitrary dependent slab distributions will not necessarily admit a convenient boundary density or a tractable time-varying unfreezing rate.
The present framework should therefore be viewed as extending sticky PDMPs to a useful class of structured priors rather than providing an automatic solution for every spike-and-slab prior.

Second, subsampling remains considerably more model specific than full-gradient evaluation.
For continuous parameters, software such as Stan can provide exact gradients of a specified log density automatically through algorithmic differentiation \citep{carpenter2017stan}.
In contrast, an efficient subsampled PDMP requires more than an unbiased gradient estimator.
The likelihood must first be decomposed into suitable factors, and the resulting stochastic event rate must admit a valid and computationally useful simulation strategy.
The OMRF construction therefore required model-specific derivations of the factor gradients and event-time bounds.
This limits the extent to which the subsampling approach can currently be treated as a generic inference method, and reflects a broader implementation challenge for practical PDMP samplers \citep{fearnhead2018piecewise, corbella2022automatic, pagani2024nuzz}.

Third, the computational comparisons with NIMBLE should not be interpreted as pure comparisons between PDMP and reversible-jump algorithms.
The samplers were implemented in different software environments and exploit different computational representations of the models.
In particular, the PDMP implementation of the RI-CLPM uses a specialized sufficient-statistic formulation, whereas the NIMBLE implementation follows a different computational route.
Differences in runtime can therefore reflect implementation choices, compiler behavior, memory access, and model-specific optimization in addition to differences between the underlying Monte Carlo algorithms.
This issue is difficult to eliminate completely in comparisons between fundamentally different sampling schemes.

More generally, comparing continuous-time PDMP samplers with discrete-time MCMC is not straightforward.
For discrete-time chains, effective sample size is conventionally defined through the autocorrelation across successive iterations \citep{geyer1992practical}.
For a PDMP, however, the process evolves continuously and arbitrarily dense observation of the same trajectory will necessarily produce highly correlated states.
Efficiency is instead naturally related to time averages over the continuous trajectory, and previous work on PDMPs has therefore expressed efficiency in terms of the amount of continuous process time or computational work required per effective sample \citep{fearnhead2018piecewise}.
Consequently, ESS per second provides a useful practical comparison, but it does not put continuous- and discrete-time samplers on a completely representation-free scale.
The results should therefore be read as comparisons between concrete implementations and workflows rather than as universal rankings of the underlying algorithms.

Two directions seem particularly useful for future work.
One is to reduce the amount of model-specific work required to construct PDMP event times.
Automatic differentiation has largely removed the need to derive gradients by hand in modern Bayesian computation, and an analogous layer for PDMPs would be valuable.
Such a system could combine automatic differentiation with numerical or symbolic bounds on event rates, automatically choosing between thinning, numerical integration, and exact cumulative hazards.
Recent work on automatic and numerical Zig--Zag implementations already moves in this direction \citep{corbella2022automatic, pagani2024nuzz}, but extending this idea to sticky boundaries and subsampled gradients would make PDMP variable selection substantially more accessible.

A second direction is to exploit the dependent-prior construction for richer forms of structured variable selection.
The model-prior term need not depend only on model size.
For network models, for example, adjacent-model probabilities could depend on latent node-specific connectivity propensities, groups of nodes, or community structure, while the slab could share scales within the same latent structure.
This would turn the prior from a generic sparsity mechanism into a direct probabilistic model for which parameters are expected to enter together.
Sticky PDMPs are particularly appealing in this setting because the required change appears locally through the adjacent-model ratio and conditional boundary density rather than requiring enumeration of the full model space.
Developing such structured priors could therefore connect PDMP variable selection with broader model-based approaches to sparsity and network organization.

\section{Conclusion}

Sticky PDMPs provide a continuous-time approach to Bayesian variable selection in which parameters can move between an exact spike at zero and a continuous slab while the remaining parameters continue to evolve.
We extended this framework to dependent model priors and dependent slab distributions, and showed how it can be combined with either sufficient-statistic gradients or unbiased stochastic gradients depending on the structure of the model.
The resulting methods performed well in the RI-CLPM simulations and demonstrated substantial computational gains from subsampling in parts of the OMRF study.

At the same time, these gains come with additional structure and implementation requirements.
Dependent priors must permit tractable boundary rates, and subsampling requires model-specific event-time constructions that are not currently supplied automatically by general probabilistic programming systems.
PDMP variable selection therefore does not replace existing MCMC methods in general, but offers a promising alternative when the geometry, prior structure, and likelihood factorization can be exploited.
For psychometric models with many potential effects or edges, these features provide a useful basis for developing more scalable and more structured approaches to Bayesian variable selection.

\bibliography{Marsman.bib, references.bib}

\newpage
\appendix
\renewcommand{\theequation}{\theappendix\arabic{equation}}
\renewcommand{\theHequation}{app.\arabic{appendix}.\arabic{equation}}

\section{Dependent Parameter Priors}
\label{app:dependent-parameter-priors}

This appendix gives the main derivations for the dependent-prior construction.
Let $A$ denote the current active set, with active coefficients $\parameters_A$, and let $j\notin A$ denote an excluded coefficient.
The state may additionally contain slab hyperparameters $\psi$.

\subsection{Boundary Rate}

We assume that the slab prior under each active set is obtained as the corresponding marginal of a common full-dimensional prior.
Specifically,
\begin{align}
p_{A+j}(\parameters_A,\parameter_j\mid\psi)=p_A(\parameters_A\mid\psi)p(\parameter_j\mid\parameters_A,\psi).
\end{align}
At the boundary $\parameter_j=0$, models $A$ and $A+j$ represent the same full parameter vector.
Their likelihood contributions and common hyperparameter priors therefore cancel, giving
\begin{align}
\frac{
\pi_{A+j}(\parameters_A,0,\psi)
}{
\pi_A(\parameters_A,\psi)
}
&=
\frac{p(A+j)}{p(A)}
p(\parameter_j=0\mid\parameters_A,\psi) \\
&=
\rho_j(A)f_j(0\mid\parameters_A,\psi).
\end{align}
The corresponding unfreezing rate is
\begin{align}
\lambda_j
=
C_v\rho_j(A)f_j(0\mid\parameters_A,\psi),
\end{align}
where $C_v=\mathbb{E}|V_j|$ is the mean absolute departure speed.

Summing over excluded coefficients gives the aggregate rate
\begin{align}
\Lambda_A=C_v\sum_{j\notin A}
\rho_j(A)f_j(0\mid\parameters_A,\psi).
\end{align}
Conditional on an aggregate unfreezing event, coefficient $j$ is selected with probability
\begin{align}
\prob{J=j\mid\text{event}}=\frac{
\rho_j(A)f_j(0\mid\parameters_A,\psi)
}{
\sum_{\ell\notin A}
\rho_\ell(A)f_\ell(0\mid\parameters_A,\psi)
}.
\end{align}

\subsection{Beta--Bernoulli Model Prior}

Suppose
\begin{align}
\rho &\sim \mathrm{Beta}(a,b), \\
\indicator_j\mid\rho &\sim \mathrm{Bernoulli}(\rho),
\qquad j=1,\ldots,d.
\end{align}
After integrating out $\rho$, an active set $A$ with $k=|A|$ has probability
\begin{align}
p(A)=\frac{B(a+k,b+d-k)}{B(a,b)}.
\end{align}
The adjacent-model ratio for adding an excluded coefficient is therefore
\begin{align}
\rho_j(A)=\frac{p(A+j)}{p(A)}=\frac{a+k}{b+d-k-1}.
\end{align}
Because this ratio depends only on the model size, it remains constant between sticky events.

\subsection{Gaussian Scale-Mixture Slabs}

Suppose that conditional on $\psi$,
\begin{align}
\parameters\mid\psi
\sim
\dnorm{\mu(\psi),\Sigma(\psi)}.
\end{align}
The mixing variables $\psi$ remain part of the PDMP state.
For $j\notin A$, Gaussian conditioning gives
\begin{align}
\parameter_j\mid\parameters_A,\psi
&\sim\dnorm{m_j,s_j^2},\\
m_j&=\mu_j+\Sigma_{jA}\Sigma_{AA}^{-1}(\parameters_A-\mu_A),\\
s_j^2&=\Sigma_{jj} - \Sigma_{jA}\Sigma_{AA}^{-1}\Sigma_{Aj}.
\end{align}
Hence,
\begin{align}
f_j(0\mid\parameters_A,\psi)=\frac{1}{s_j}\phi\left(\frac{m_j}{s_j}\right),
\end{align}
and
\begin{align}
\Lambda_A=C_v\sum_{j\notin A}\rho_j(A)\frac{1}{s_j}\phi\left(\frac{m_j}{s_j}\right).
\end{align}

For an exchangeable Gaussian slab,
\begin{align}
\mu(\psi)&=\mu\mathbf{1}_d, \\
\Sigma(\psi)&= uI_d+v\mathbf{1}_d\mathbf{1}_d^\top,
\end{align}
all excluded coefficients have the same conditional distribution.
Writing $k=|A|$,
\begin{align}
m_A&=\mu+
\frac{v}{u+kv}
\sum_{i\in A}(\parameter_i-\mu),\\
s_k^2 &=\frac{u\bigl(u+(k+1)v\bigr)}{u+kv}.
\end{align}
The aggregate rate therefore reduces to
\begin{align}
\Lambda_A=C_v\frac{1}{s_k}\phi\left(\frac{m_A}{s_k}\right)\sum_{j\notin A}\rho_j(A).
\end{align}

\subsection{Simulation of the Unfreezing Clock}

Along the deterministic PDMP trajectory, the aggregate rate becomes
\begin{align}
\Lambda_A(t)
=C_v\sum_{j\notin A}\rho_j(A)
f_j\bigl(0\mid\parameters_A(t),\psi(t)\bigr).
\end{align}
The next unfreezing time satisfies
\begin{align}
\int_0^T \Lambda_A(t)\,\mathrm{d}t = E,
\qquad
E\sim\mathrm{Exponential}(1).
\end{align}

For Zig--Zag and Bouncy Particle dynamics, the active coefficients move linearly between events.
Under a fixed zero-centered exchangeable Gaussian slab,
\begin{align}
m_A(t)=a+bt,
\end{align}
while $s_k$ is constant.
Writing
\begin{align}
W_A=C_v\sum_{j\notin A}\rho_j(A),
\end{align}
the cumulative hazard is
\begin{align}
H_A(T)=\frac{W_A}{|b|}
\left|\Phi\left(\frac{a+bT}{s_k}\right) - \Phi\left(\frac{a}{s_k}\right)\right|.
\end{align}
for $b\neq0$.
If $b=0$, the rate is constant.

For Boomerang dynamics,
\begin{align}
m_A(t)=a+B\cos(t-\delta),
\end{align}
and hence
\begin{align}
\Lambda_A(t)=W_A
\frac{1}{s_k}
\phi\left(
\frac{a+B\cos(t-\delta)}{s_k}
\right).
\end{align}
This one-dimensional periodic rate is simulated by thinning using a piecewise upper envelope.

\section{OMRF Subsampling and Event-Time Derivations}
\label{app:omrf-clocks}

This appendix gives the OMRF-specific likelihood gradients and subsampling construction.

\subsection{OMRF Pseudolikelihood Gradients}

Let $x_n=(x_{n1},\ldots,x_{nP})$ denote the response vector for person $n$.
For node $i$, define
\begin{align}
\eta_{ni}=\sum_{j\neq i}x_{nj}\sigma_{ij},
\end{align}
and
\begin{align}
q_{niu}=\frac{
\exp{\alpha_{iu}+u\eta_{ni}}
}{
\sum_{r=0}^{K_i-1}
\exp{\alpha_{ir}+r\eta_{ni}}
}.
\end{align}
The person--node log-pseudolikelihood contribution is
\begin{align}
\ell_{ni}(\parameters)
=
\alpha_{i,x_{ni}}
+
x_{ni}\eta_{ni}
-
\log
\sum_{u=0}^{K_i-1}
\exp{\alpha_{iu}+u\eta_{ni}}.
\end{align}

Writing
\begin{align}
\bar{x}_{ni}
=
\sum_{u=0}^{K_i-1}u q_{niu},
\end{align}
the threshold gradients are
\begin{align}
\frac{\partial \ell_{ni}}{\partial \alpha_{ir}}
=
\mathbb{1}(x_{ni}=r)-q_{nir},
\end{align}
and
\begin{align}
\frac{\partial \ell_{ni}}{\partial \eta_{ni}}
=
x_{ni}-\bar{x}_{ni}.
\end{align}
For an edge $(j,k)$, the contribution of person $n$ is
\begin{align}
\frac{\partial \ell_n}{\partial \sigma_{jk}}
=
x_{nk}(x_{nj}-\bar{x}_{nj})
+
x_{nj}(x_{nk}-\bar{x}_{nk}).
\end{align}
The likelihood contribution to the negative log-posterior gradient is the negative of these expressions.

\subsection{Subsampled Gradient}

Index the $L=NP$ person--node factors by $a=(n,i)$ and define
\begin{align}
g_a(\parameters)
=
-\nabla\ell_a(\parameters).
\end{align}
Then
\begin{align}
\nabla U(\parameters)
=
\nabla U_{\mathrm{prior}}(\parameters)
+
\sum_{a=1}^{L}g_a(\parameters).
\end{align}
If $J_1,\ldots,J_m$ are sampled independently with probabilities $\pi_a$, an unbiased estimator is
\begin{align}
\widehat{\nabla U}(\parameters)
=
\nabla U_{\mathrm{prior}}(\parameters)
+
\frac{1}{m}
\sum_{b=1}^{m}
\frac{g_{J_b}(\parameters)}{\pi_{J_b}}.
\end{align}
Indeed,
\begin{align}
\mathbb{E}\left[
\frac{g_{J_b}(\parameters)}{\pi_{J_b}}
\right]
=
\sum_{a=1}^{L}g_a(\parameters).
\end{align}
Under uniform sampling, $\pi_a=1/L$ and
\begin{align}
\widehat{\nabla U}(\parameters)
=
\nabla U_{\mathrm{prior}}(\parameters)
+
\frac{L}{m}
\sum_{b=1}^{m}
g_{J_b}(\parameters).
\end{align}
The prior gradient is always evaluated exactly.

\subsection{Stochastic Event Rates}

For Zig--Zag coordinate $i$, let
\begin{align}
\widehat{\lambda}_i(\parameters,\velocities)
=
\left[
\velocity_i\widehat{g}_i(\parameters)
\right]^+.
\end{align}
If $F_i$ flips the sign of the $i$th velocity coordinate, then
\begin{align}
\widehat{\lambda}_i(\parameters,\velocities)
-
\widehat{\lambda}_i(\parameters,F_i\velocities)
&=
\velocity_i\widehat{g}_i(\parameters).
\end{align}
Taking expectations gives
\begin{align}
\mathbb{E}\left[
\widehat{\lambda}_i(\parameters,\velocities)
-
\widehat{\lambda}_i(\parameters,F_i\velocities)
\right]
=\velocity_i
\frac{\partial U(\parameters)}{\partial\parameter_i},
\end{align}
which is the rate-difference identity required for the Zig--Zag invariant distribution.

The same principle is used for the Bouncy Particle and Boomerang samplers: the stochastic event mechanism is constructed so that its average rate satisfies the corresponding balance identity.
Provided the event-time construction is valid, the resulting sampler targets the OMRF pseudoposterior exactly.
This does not remove the approximation introduced by replacing the full OMRF likelihood with the pseudolikelihood.

\end{document}